\documentclass{appolb}
\usepackage{array}
\usepackage{graphicx}

\begin{document}
\title{Scalar mesons: fifty years of challenging the quark model\thanks
{Presented by G.~Rupp at Workshop ``Excited QCD 2018'',
Kopaonik, Serbia, March 11--15, 2018.}}
\author{George Rupp\address{Centro de F\'{\i}sica e Engenharia de Materiais
Avan\c{c}ados, Instituto Superior T\'{e}cnico, Universidade de Lisboa,
P-1049-001, Portugal}
\\[5mm]
Eef van Beveren\address{Centro de F\'{\i}sica da UC, Departamento de
F\'{\i}sica, Universidade de Coimbra, P-3004-516, Portugal}
}
\maketitle

\begin{abstract}
Half a century of work on the light scalar mesons $f_0(500)$,
$f_0(980)$, $K_0^\star(700)$, and $a_0(980)$ is briefly reviewed. After
summarising all light scalar candidates in the Review of Particle Physics
since 1963, a selection of different theoretical and phenomenological
descriptions is presented, including pure meson-meson models, a tetraquark
construction, unitarised quark-meson models, unitarised effective chiral
approaches, and a very recent lattice-QCD simulation. 
\end{abstract}
The light scalar-meson nonet $f_0(500)$ (alias $\sigma$), $f_0(980)$,
$K_0^\star(700)$ (alias $\kappa$), and $a_0(980)$ \cite{PDG18} has been vexing
experimentalists as well as theorists for more than fifty years by now.
Especially the isoscalar $\sigma$ and isodoublet $\kappa$ in $S$-wave
$\pi\pi$ and $K\pi$ scattering, respectively, have for many years been
considered doubtful as genuine resonances. Only very recently \cite{PDG18}
the $\kappa$ meson was included in the PDG Meson Summary Table of the
Review of Particle Physics (RPP). But also
the isoscalar $f_0(980)$ and isovector $a_0(980)$ have for several years been
questioned. In Table~\ref{PDG} we list all light scalars that have appeared in
the PDG tables since the earliest RPP days, including states with uncertain
$J^{PC}$ or even initially identified as having $J\geq1$. The reported
masses and widths or pole positions, when available, can be found in the 
corresponding references. Only those PDG editions are included in
the table that report any change of one or more 
\begin{table}[!h]
\caption{The light scalar-meson nonet in \em The Review of Particle Physics \em
\/(RPP). $^\dag J^{PC}=0^{++}$?, $^\ddag J^{PC}$ unknown,
$^{\dag\dag} J^{PC}=(\mbox{Even})^{++}$, $^{\ddag\ddag} J\geq1$;
$J^{PC}=0^{++}$ otherwise.}
\begin{tabular}{l||c|c|c|c}
\hline\hline
Year & $f_0(500)$ & $K_0^\star(700)$ & $f_0(980)$ & $a_0(980)$ \\ \hline
1963\cite{PDG63-1}&$\omega_{ABC}(317)^{\ddag}$&$K_1^{**}(730)^{\ddag\ddag}$
                  & $\kappa_2(1040)^{\dag\dag}$ & $\chi_1(1045)^\ddag$ \\
1963\cite{PDG63-2}& $\omega_{ABC}(317)$ & $\kappa(725)^{\ddag\ddag}$
                           & ?$(1040)^{\dag\dag}$ & ?$^-(1000)^\ddag$ \\
1964\cite{PDG64}  & $\sigma(390)^\dag$ & $\kappa(725)^\ddag$ 
                  & $K_1K_1(1020)^{\dag\dag}$ & - \\
1965\cite{PDG65}  & \{$\begin{array}{ll}\!\!\!\sigma(390)^\dag\\[-0.5mm]
                  \!\!\!S0(\pi\pi,700)\end{array}$ & $\kappa(725)^\ddag$
                  & $K_1K_1(1020)^{\dag\dag}$ & - \\
1966\cite{PDG66}  & \{$\begin{array}{ll}\!\!\!\sigma(390)^\dag\\[-0.5mm]
                  \!\!\!S^0(720)\end{array}$ & $\kappa(725)^\ddag$
                  & $K\bar{K}_0(1068)$ & - \\
1967\cite{PDG67}  & \{$\begin{array}{ll}\!\!\!\sigma(410) \\[-0.5mm]
                  \!\!\!\epsilon(700)\end{array}$ & $\kappa(725)^\ddag$
                  & $\eta_V(1050)$ & $\delta(965)^\ddag$ \\
1968\cite{PDG68}  & \{$\begin{array}{ll}\!\!\!\sigma(410) \\[-0.5mm]
                  \!\!\!\epsilon(730)\end{array}$ & $\kappa(725)^\ddag$
   & $\eta_V(1070)$ & \{$\begin{array}{ll}\!\!\!\delta(963)^\ddag \\[-0.5mm]
                                          \!\!\!\pi_N(1016)\end{array}$ \\
1969\cite{PDG69}  & \{$\begin{array}{ll}\!\!\!\sigma(410) \\[-0.5mm]
                  \!\!\!\eta_{0+}(720)\end{array}$ & $\kappa(725)^\ddag$
 & $\eta_{0^+}(1070)$ & \{$\begin{array}{ll}\!\!\!\delta(963)^\ddag \\[-0.5mm]
                                          \!\!\!\pi_N(1016)\end{array}$ \\
1970\cite{PDG70-1} & \{$\begin{array}{ll}\!\!\!\sigma(410) \\[-0.5mm]
                  \!\!\!\eta_{0+}(700)\end{array}$ & $\kappa(725)^\ddag$
 & $\eta_{0+}(1060)$ & \{$\begin{array}{ll}\!\!\!\delta(962)^\ddag \\[-0.5mm]
                                          \!\!\!\pi_N(1016)\end{array}$ \\
1970\cite{PDG70-2} & \{$\begin{array}{ll}\!\!\!\sigma(410) \\[-0.5mm]
            \!\!\!\eta_{0+}/\epsilon(700)\end{array}$ & $\kappa(725)^\ddag$
&$\eta_{0+}/S^\star(1060)$&\{$\begin{array}{ll}\!\!\!\delta(966)^\ddag\\[-0.5mm]
                                          \!\!\!\pi_N(1016)\end{array}$ \\
1971\cite{PDG71} & \{$\begin{array}{ll}\!\!\!\sigma(410) \\[-0.5mm]
\!\!\!\eta_{0+}/\epsilon\mbox{(700--1000)}\end{array}$ & $\kappa(725)^\ddag$
&$\eta_{0+}/S^\star(1070)$&\{$\begin{array}{ll}\!\!\!\delta(962)^\ddag\\[-0.5mm]
                                          \!\!\!\pi_N(1016)\end{array}$ \\
1972\cite{PDG72}  & $\epsilon$ & $\kappa$ 
                  & $S^\star(1000)$ & $\pi_N(975)$ \\
1973\cite{PDG73}  & $\epsilon$ & $\kappa$ 
                  & $S^\star(997)$ & $\delta(970)$ \\
1974\cite{PDG74}  & $\epsilon$ & $\kappa$ 
                  & $S^\star(993)$ & $\delta(970)$ \\
1976\cite{PDG76}  & $\epsilon(1200)$ & $\kappa(1250)$ 
                  & $S^\star(993)$ & $\delta(970)$ \\
1978\cite{PDG78}  & $\epsilon(1300)$ & $\kappa(1400)$ 
                  & $S^\star(980)$ & $\delta(980)$ \\
1980\cite{PDG80}  & $\epsilon(1300)$ & $\kappa(1500)$ 
                  & $S^\star(980)$ & $\delta(980)$ \\
1982\cite{PDG82}  & $\epsilon(1300)$ & $\kappa(1350)$ 
                  & $S^\star(975)$ & $\delta(980)$ \\
1984\cite{PDG84}  & $\epsilon(1300)$ & $\kappa(1350)$ 
                  & $S/S^\star(975)$ & $\delta(980)$ \\
1986\cite{PDG86}  & $f_0(1300)$ & $K_0^\star(1350)$ 
                  & $f_0(975)$ & $a_0(980)$ \\
1988\cite{PDG88}  & $f_0(1400)$ & $K_0^\star(1430)$ 
                  & $f_0(975)$ & $a_0(980)$ \\
1990\cite{PDG90}  & $f_0(1400)$ & $K_0^\star(1350)$ 
                  & $f_0(975)$ & $a_0(980)$ \\
1992\cite{PDG92}  & $f_0(1400)$ & $K_0^\star(1430)$ 
                  & $f_0(975)$ & $a_0(980)$ \\
1994\cite{PDG94}  & $f_0(1300)$ & $K_0^\star(1430)$ 
                  & $f_0(980)$ & $a_0(980)$ \\
1996\cite{PDG96}  & $f_0\mbox{(400--1200)}$ & $K_0^\star(1430)$ 
                  & $f_0(980)$ & $a_0(980)$ \\
2002\cite{PDG02}  & $f_0(600)$ & $K_0^\star(1430)$ 
                  & $f_0(980)$ & $a_0(980)$ \\
2004\cite{PDG04}  & $f_0(600)$ & $K_0^\star(800)$ 
                  & $f_0(980)$ & $a_0(980)$ \\
2012\cite{PDG12}  & $f_0(500)$ & $K_0^\star(800)$ 
                  & $f_0(980)$ & $a_0(980)$  \\
2018\cite{PDG18}  & $f_0(500)$ & $K_0^\star(700)$ 
                  & $f_0(980)$ & $a_0(980)$ \\[-5mm]
\end{tabular}
\label{PDG}
\end{table}
scalar entries with respect to the previous edition.

The difficulty in extracting the light scalar mesons from the data originates
in Adler zeros in the associated $S$-wave amplitudes, overlapping resonances,
and strong inelasticities. In the remainder of this brief review we shall
present a selection of theoretical and phenomenological modellings of the light
scalars, as well as a very recent lattice calculation of the $\sigma$ meson.
For a much more comprehensive list of $\sigma$ and $\kappa$
works, see Ref.~\cite{Eef}. \\[2mm]
{\textbf{1) \textit{``Yang-Mills fields and pseudoscalar meson
scattering''},
D.\ Ia\-golnitzer, J.~Zinn-Justin, J.~B.~Zuber \cite{NPB60p233}} \\[0.5mm]
A Lagrangian model for the scattering of pseudoscalar mesons is formulated,
with exchanges of the vector mesons $\rho$, $K^\star$, and $\phi$. The model
is not renormalisable, so subtraction constants are used. Unitarisation is
done applying the Pad\'{e} method to Born plus one-loop diagrams. Fits to the
data then yield, for $S$-wave scattering, the following scalar masses,
widths (in MeV):
$\epsilon(460,675)$; $S^\star(990,40)$;
$\kappa(665,840)$; $\pi_N(775,610)$ (also see Table 1). \\[2mm]
{\textbf{2) \textit{``Multiquark hadrons I. Phenomenology of 
\boldmath$Q^2\bar{Q}^2$ mesons''},
R.\ L.\ Jaffe \cite{PRD15p267}} \\[0.5mm]
The light scalars are modelled as $q^2\bar{q}^2$ states in the MIT Bag Model.
A very large attractive colour-hyperfine interaction results in the following
low yet purely real masses (in MeV): $\epsilon(700)$; $S^\star(1100)$; 
$\kappa(900)$; $\delta(1100)$. \\[2mm]
{\textbf{3) \textit{``Dynamical Symmetry Breaking and the Sigma-Meson Mass in
Quantum Chromodynamics''},
R.\ Delbourgo, M.\ D.\ Scadron \cite{PRL48p379}} \\[0.5mm]
The spontaneous breakdown of chiral symmetry is analysed dynamically via
bound-state Bethe-Salpeter equations. While in general spontaneous mass
generation is linked to a massless pseudoscalar pion and to no specific 
constraint on a massive scalar meson, for the particular theory of
asymptotically free QCD it is shown that a scalar $\sigma$ meson should
exist with mass $m_\sigma\approx$ 600--700 MeV. \\[2mm]
{\textbf{4) \textit{``A low lying scalar meson nonet in a unitarized meson
model''},
E.~van Beveren \textit{et al.} \cite{ZPC30p615}} \\[0.5mm]
A unitarised quark model previously fitted to light and heavy pseudoscalar
and vector mesons is applied to scalar mesons as normal $P$-wave $q\bar{q}$
states, with unaltered parameters. Apart from the standard scalars between
1.3 and 1.5 GeV, an additional nonet below 1 GeV is dynamically generated,
with the following resonance pole positions (in MeV): \\
$\epsilon(470-i208)$, $S^\star(994-i20)$, $\kappa(727-i263)$,
$\delta(968-i28)$. \\
These values are still compatible with present-day PDG limits. \\[2mm]
{\textbf{5) \textit{``Relativistic effects in scalar meson dynamics''},
R.~Kaminski, L.~Lesniak, J.~P.~Maillet \cite{PRD50p3145}} \\[0.5mm]
A purely mesonic model for the coupled $S$-wave channels $\pi\pi$ and
$K\bar{K}$, with phenomenological separable potentials, is solved through a
relativistic Lippmann-Schwinger equation. Fitting the parameters to data
yields the following masses, widths (in MeV): $f_0(500)(506,494)$;
$f_0(980)(973,29)$. \\[2mm]
{\textbf{6) \textit{``Structure of the scalar mesons \boldmath$f_0(980)$ and
\boldmath$a_0(980)$''},
G.\ Jans\-sen, B.~C.~Pearce, K.~Holinde, J.~Speth \cite{PRD52p2690}} \\[0.5mm]
A Blankenbecler-Sugar equation is solved for $S$-wave $\pi\pi,K\bar{K}$, and
$\pi\eta$ scattering, with $t$-channel vector-meson exchanges and $s$-channel
scalar exchanges. Fits to the data result in the following poles (in MeV): \\
$\sigma(387-i305)$; $f_0(980)(1015-i15)$; $a_0(980)(991-i101)$.
\\[2mm]
{\textbf{7) \textit{``Confirmation of the Sigma Meson''},
N.~A.~Tornqvist, M.\ Roos \cite{PRL76p1575}}} \\[0.5mm]
Bare scalar $q\bar{q}$ states are coupled to channels of two pseudoscalar
mesons, in a unitarised quark-meson model. Fitting the data predicts the
following scalar-meson poles (in MeV): $\sigma(470-i250)$; 
$f_0(980)(1006-i17)$; $a_0(980)(1094-i145)$. Note: no $\kappa$ pole found.
\\[2mm]
{\textbf{8) \textit{``Simple description of \boldmath$\pi\pi$ scattering
to 1 GeV''},
M.~Harada, F.~Sannino, J.~Schechter \cite{PRD54p1991}}} \\[0.5mm]
$\pi\pi$ amplitudes from an effective nonlocal chiral Lagrangian are written
as a sum of a relativistic Breit-Wigner form plus a non-resonant background,
with local unitarity and crossing symmetry satisfied. Fits produce
the following scalar masses, widths (in MeV): $\sigma(559,370)$;
$f_0(980)(980,\mbox{40--400})$. \\[2mm]
{\textbf{9) \textit{``An Analysis of \boldmath$\pi\pi$-Scattering Phase Shift
and Existence of \boldmath$\sigma(555)$ Particle''},
S.~Ishida \textit{et al.} \cite{PTP95p745}} \\[0.5mm]
The unitary interfering-amplitude method is employed to describe $S$-wave
$\pi\pi$ scattering, including a negative background phase instead of the
usual Adler zero. For $f_0(980)$, the $\pi\pi\to K\bar{K}$ data are
included, too. Fits give the following masses, widths (in MeV):
$\sigma(553,243)$; $f_0(980)(993,\sim\!100)$. \\[2mm]
{\textbf{10) \textit{``Meson-meson interactions in a nonperturbative chiral
approach''},
J.~A.\ Oller, E.~Oset, J.~R.~Pelaez \cite{PRD59p074001}} \\[0.5mm]
Amplitudes from $\mathcal{O}(p^2)$ and $\mathcal{O}(p^4)$ chiral
Lagrangians are unitarised with the inverse-amplitude method. Fits to the
$S$-wave data yield the following scalar-meson poles (in MeV): \\
$\sigma(442-i272)$; $f_0(980)(994-i14)$; $\kappa(770-i250)$;
$a_0(980)(1055-i21)$. \\[2mm]
{\textbf{11) \textit{``Comments on the \boldmath$\sigma$ and
\boldmath$\kappa$''},
D.~V.\ Bugg \cite{PLB572p1}} \\[0.5mm]
Relativistic Breit-Wigner forms with Adler zeros included in the
energy-dependent widths are used to fit combined $S$-wave data
from elastic scattering and production processes for $\pi\pi$, and
elastic data only for $K\pi$. Resulting poles of $\sigma$ and
$\kappa$ (in MeV): $(533\pm25)-i(249\pm25)$, $(722\pm60)-i(386\pm50)$. \\[2mm]
{\textbf{12) \textit{``Mass and width of the lowest resonance in QCD''},
I.\ Caprini, G.\ Colangelo, H.\ Leutwyler \cite{PRL96p132001}} \\[0.5mm]
A twice-subtracted dispersion relation (Roy equation) is employed to extract
a $\sigma$ pole position from $\pi\pi$ data, while fixing the subtraction
constants via chiral perturbation theory.
Result: $(441^{+16}_{-8}-i272^{+9}_{-12.5})$ MeV. \\[2mm]
{\textbf{13) \textit{``Isoscalar \boldmath$\pi\pi$ Scattering and the
\boldmath$\sigma$ Meson Resonance from QCD''},
R.\ A.\ Briceno, J.~J.\ Dudek, R.~G.\ Edwards, D.~J.\ Wilson
\cite{PRL118p022002}} \\[0.5mm]
A lattice calculation of energy-dependent $S$-wave isoscalar $\pi\pi$ phase
shifts is carried out satisfying elastic unitarity, with both $q\bar{q}$ and
$\pi\pi$ interpolators included, for $\pi$ masses of 391 and 236 MeV. A
$\pi\pi$ bound state shows up in the former case and a broad resonance in the
latter. This resonance resembles the $\sigma$ meson, though its precise pole
position depends on the employed parametrisation. Future work will
aim at using even lighter $u,d$ quarks and pions, besides imposing constraints
from causality and crossing symmetry. \\

To conclude, in this minireview we have summarised all RPP entries of light
scalar-meson candidates, starting from the earliest available data. Moreover,
a selection of typical and innovative model approaches has been provided, which 
we hope sheds some more light on the nature of these enigmatic mesons. Finally,
a very recent lattice calculation appears to support their interpretation as
strongly unitarised $q\bar{q}$ states, as pioneered in Ref.~\cite{ZPC30p615}.


\begin{thebibliography}{99}

\bibitem{PDG18}
M.~Tanabashi {\it et al.} [Particle Data Group],
{\it Phys.\ Rev.\ D} \/{\bf 98}, 030001 (2018).

\bibitem{PDG63-1} 
M.~Roos,
{\it Rev.\ Mod.\ Phys.} {\bf 35}, 314 (1963).

\bibitem{PDG63-2} 
M.~Roos,
{\it Nucl.\ Phys.} {\bf 52}, 1 (1964).

\bibitem{PDG64} 
A.~H.~Rosenfeld {\it et al.},
{\it Rev.\ Mod.\ Phys.} {\bf 36}, 977 (1964).

\bibitem{PDG65} 
A.~H.~Rosenfeld {\it et al.},
{\it Rev.\ Mod.\ Phys.} {\bf 37}, 633 (1965).

\bibitem{PDG66} 
A.~H.~Rosenfeld {\it et al.}, \em
``Mesons'', \em
in Procs.\ of ``8th Brandeis University Summer Institute in Theoretical
Physics: Particle symmetries and axiomatic field theory'', 
Waltham, MA, USA, 1965,
{\tt http://pdg.lbl.gov/rpp-archive/files/cm-p00040259.pdf}

\bibitem{PDG67} 
A.~H.~Rosenfeld {\it et al.}, 
{\it Rev.\ Mod.\ Phys.} {\bf 39}, 1 (1967).

\bibitem{PDG68} 
A.~H.~Rosenfeld {\it et al.}, 
{\it Rev.\ Mod.\ Phys.} {\bf 40}, 77 (1968).

\bibitem{PDG69} 
N.~Barash-Schmidt {\it et al.},
{\it Rev.\ Mod.\ Phys.} {\bf 41}, 109 (1969).

\bibitem{PDG70-1} 
A.~Barbaro-Galtieri {\it et al.},
{\it Rev.\ Mod.\ Phys.} {\bf 42}, 87 (1970).

\bibitem{PDG70-2}
[Particle Data Group],
{\it Phys.\ Lett.} {\bf 33B}, 1 (1970).

\bibitem{PDG71} 
A.~Rittenberg {\it et al.},
{\it Rev.\ Mod.\ Phys.} {\bf 43}, S1 (1971).

\bibitem{PDG72} 
P.~Soeding {\it et al.},
{\it Phys.\ Lett.} {\bf 39B}, 1 (1972).

\bibitem{PDG73} 
T.~A.~Lasinski {\it et al.},
{\it Rev.\ Mod.\ Phys.} {\bf 45}, S1 (1973).

\bibitem{PDG74} 
V.~Chaloupka {\it et al.} \/[Particle Data Group],
{\it Phys.\ Lett.} {\bf 50B}, 1 (1974).
 
\bibitem{PDG76} 
T.~G.~Trippe {\it et al.} [Particle Data Group],
{\it Rev.\ Mod.\ Phys.} {\bf 48}, S1 (1976)
[{\it Erratum-ibid} \/{\bf 48}, 497 (1976)].

\bibitem{PDG78} 
C.~Bricman {\it et al.} [Particle Data Group],
{\it Phys.\ Lett.} {\bf 75B}, 1 (1978).

\bibitem{PDG80} 
R.~L.~Kelly {\it et al.} [Particle Data Group],
{\it Rev.\ Mod.\ Phys.} {\bf 52}, S1 (1980).

\bibitem{PDG82} 
M.~Roos {\it et al.} [Particle Data Group],
{\it Phys.\ Lett.} {\bf 111B}, 1 (1982).

\bibitem{PDG84} 
C.~G.~Wohl {\it et al.} [Particle Data Group],
{\it Rev.\ Mod.\ Phys.} {\bf 56}, S1 (1984).

\bibitem{PDG86} 
M.~Aguilar-Benitez {\it et al.} [Particle Data Group],
{\it Phys.\ Lett.} {\bf 170B}, 1 (1986).

\bibitem{PDG88} 
G.~P.~Yost {\it et al.} [Particle Data Group],
{\it Phys.\ Lett.\ B} \/{\bf 204}, 1 (1988).

\bibitem{PDG90} 
J.~J.~Hernandez {\it et al.} [Particle Data Group],
{\it Phys.\ Lett.\ B} \/{\bf 239}, 1 (1990)
[{\it Erratum-ibid} \/{\bf 253}, 524 (1991)].

\bibitem{PDG92} 
K.~Hikasa {\it et al.} [Particle Data Group],
{\it Phys.\ Rev.\ D} \/{\bf 45}, S1 (1992)
[{\it Erratum-ibid} \/{\bf 46}, 5210 (1992)].

\bibitem{PDG94} 
L.~Montanet {\it et al.} [Particle Data Group],
{\it Phys.\ Rev.\ D} \/{\bf 50}, 1173 (1994).

\bibitem{PDG96} 
R.~M.~Barnett {\it et al.} [Particle Data Group],
{\it Phys.\ Rev.\ D} \/{\bf 54}, 1 (1996).

\bibitem{PDG02} 
K.~Hagiwara {\it et al.} [Particle Data Group],
{\it Phys.\ Rev.\ D} \/{\bf 66}, 010001 (2002).

\bibitem{PDG04} 
S.~Eidelman {\it et al.} [Particle Data Group],
{\it Phys.\ Lett.\ B} \/{\bf 592}, 1 (2004).

\bibitem{PDG12} 
J.~Beringer {\it et al.} [Particle Data Group],
{\it Phys.\ Rev.\ D} \/{\bf 86}, 010001 (2012).

\bibitem{Eef}
Eef van Beveren,
{\tt http://cft.fis.uc.pt/eef/sigkap0.htm}

\bibitem{NPB60p233}
D.~Iagolnitzer, J.~Zinn-Justin, J.~B.~Zuber,
{\it Nucl.\ Phys.\ B} \/{\bf 60}, 233 (1973).

\bibitem{PRD15p267} 
R.~L.~Jaffe,
{\it Phys.\ Rev.\ D} \/{\bf 15}, 267 (1977).

\bibitem{PRL48p379} 
R.~Delbourgo and M.~D.~Scadron,
{\it Phys.\ Rev.\ Lett.} {\bf 48}, 379 (1982).

\bibitem{ZPC30p615} 
E.~van Beveren {\it et al.},
{\it Z.\ Phys.\ C} \/{\bf 30}, 615 (1986)
[arXiv:0710.4067 [hep-ph]].

\bibitem{PRD50p3145} 
R.~Kaminski, L.~Lesniak, J.~P.~Maillet,
{\it Phys.\ Rev.\ D} \/{\bf 50}, 3145 (1994)
[arXiv:hep-ph/9403264].

\bibitem{PRD52p2690} 
G.~Janssen, B.~C.~Pearce, K.~Holinde, J.~Speth,
{\it Phys.\ Rev.\ D} \/{\bf 52}, 2690 (1995)
[nucl-th/9411021].

\bibitem{PRL76p1575} 
N.~A.~Tornqvist, M.~Roos,
{Phys.\ Rev.\ Lett.} {\bf 76}, 1575 (1996)
[arXiv:hep-ph/9511210].

\bibitem{PRD54p1991} 
M.~Harada, F.~Sannino, J.~Schechter,
{\it Phys.\ Rev.\ D} \/{\bf 54}, 1991 (1996)
[arXiv:hep-ph/9511335].

\bibitem{PTP95p745} 
S.~Ishida {\it et al.},
{\it Prog.\ Theor.\ Phys.} {\bf 95}, 745 (1996)
[arXiv:hep-ph/9610325].

\bibitem{PRD59p074001} 
J.~A.~Oller, E.~Oset, J.~R.~Pelaez,
{\it Phys.\ Rev.\ D} \/{\bf 59}, 074001 (1999)
[{\it Errata-ibid} {\bf 60}, 099906 (1999);
                   {\bf 75}, 099903 (2007)]
[arXiv:hep-ph/9804209].

\bibitem{PLB572p1} 
D.~V.~Bugg,
{\it Phys.\ Lett.\ B} \/{\bf 572}, 1 (2003)
[{\it Erratum-ibid} \/{\bf 595}, 556 (2004)].

\bibitem{PRL96p132001} 
I.~Caprini, G.~Colangelo, H.~Leutwyler,
{\it Phys.\ Rev.\ Lett.} \/{\bf 96}, 132001 (2006)
[arXiv:hep-ph/0512364].

\bibitem{PRL118p022002} 
R.~A.~Briceno, J.~J.~Dudek, R.~G.~Edwards, D.~J.~Wilson,
{\it Phys.\ Rev.\ Lett.} \/{\bf 118}, 022002 (2017)
[arXiv:1607.05900 [hep-ph]].

\end{thebibliography}
\end{document}